\documentclass[10pt,conference]{IEEEtran}
\usepackage{cite}
\usepackage{amsmath,amssymb,amsfonts}
\usepackage{algorithmic}
\usepackage{graphicx}
\usepackage{textcomp}
\usepackage{xcolor}
\usepackage{hyperref}
\def\BibTeX{{\rm B\kern-.05em{\sc i\kern-.025em b}\kern-.08em
    T\kern-.1667em\lower.7ex\hbox{E}\kern-.125emX}}
\begin{document}

\title{Systems Modeling for novice engineers to comprehend software products better}

\author{
\IEEEauthorblockN{Mrityunjay Kumar}
\IEEEauthorblockA{\textit{Software Engineering Research Center} \\
IIIT Hyderabad, India\\
mrityunjay.k@research.iiit.ac.in \\
}

\and

\IEEEauthorblockN{Venkatesh Choppella}
\IEEEauthorblockA{\textit{Software Engineering Research Center} \\
IIIT Hyderabad, India\\
venkatesh.choppella@iiit.ac.in \\
}

}
\maketitle

\begin{abstract}
  One of the key challenges for a novice engineer in a product company is to comprehend the product sufficiently and quickly. It can take anywhere from six months to several years for them to attain mastery but they need to start delivering results much before. SaaS (Software-as-a-Service) products have sophisticated system architecture which adds to the time and effort of understanding them. On the other hand, time available to new hires for product understanding continues to be short and getting shorter, given the pressure to deliver more in less time.   

Constructivist theory views learning as a personal process in which the learner constructs new knowledge for themselves. Building and refining a mental model is the key way in which they learn, similar to how the brain operates. This paper presents an approach to improve system comprehension process by using a system model that a) acts as a transitional object to aid and refine the mental model of the learner, and b) captures the current understanding of the dynamics of the software system in a way that can be reasoned with and simulated. 

We have adapted discrete systems modeling techniques and used a transition system as a lightweight modeling language. Such a model can be used by novice engineers during their product ramp-up phase to build a model of the software system that captures their knowledge of the system and aid their mental model. The paper also presents a learning approach in which the learners create and refine these models iteratively using the available and newly uncovered knowledge about the software system. We hypothesize that by leveraging this modeling language and approach, novice engineers can reduce the time it takes them to achieve desired proficiency level of system comprehension.   

This paper presents early ideas on this language and approach. Future work will include designing a course to teach this language and approach to final-year students and novice engineers so that this hypothesis can be tested. 

\end{abstract}

\begin{IEEEkeywords}
System Modeling, SaaS Product, Novice Engineer, Mental Model, Software System Comprehension
\end{IEEEkeywords}

\section{Introduction}
\label{sec:introduction}
\textit{Software system comprehension}: Comprehension can mean several things. In this paper, we are focused on understanding of the dynamics of the system since they are key to systems understanding. We define software system comprehension to be an understanding of the system behavior on the whole, how parts of the system behave, and how they interconnect to deliver the observable whole system behavior. 

\textit{Comprehending software systems take time}. New engineering hires in most successful product companies need to spend significant time in understanding the product. Sim et al. \cite{sim1998ramp} as well as Zhou and Mockus \cite{10.1145/1882291.1882313} have pointed to long time it takes the engineers to become fluent on the product, almost three years in some cases. Our interviews with hiring managers at SaaS product startups suggest it takes their new hires six to twelve months to get sufficient proficiency of the product. Most product companies, including large software companies like Microsoft \cite{ju2021case}, Google \cite{johnson2010learning} and IBM \cite{dagenais2010moving} have a similar onboarding process; they provide a few weeks of onboarding time to understand the product and then assign active development work to them. The time provided is rarely enough for sufficient comprehension. A novice engineer in a SaaS product company finds it extra hard since they have very little prior exposure to large systems. 

\textit{Growth of SaaS and Cloud products makes comprehension more time-consuming}. Number of Small and Medium Businesses (SMBs) spending more than a million USD a month on Cloud went up from 38\% to 53\% (\href{https://info.flexera.com/CM-REPORT-State-of-the-Cloud}{Flexera 2022 State of the Cloud Report}). Enterprises are migrating to cloud and purchasing more SaaS \cite{10.1145/1721654.1721672} products and SaaS applications make up almost 70\% of total company software use (\href{https://pages.bettercloud.com/rs/719-KZY-706/images/2020_StateofSaaSOpsReport.pdf}{BetterCloud 2020 State of SaaSOps}). A typical SaaS product is a distributed system using multiple external APIs and a service-oriented architecture. Multi-tenancy (multiple customers are served using same software system and code) mean that small defects can have significant impact across the customer base and this has driven SaaS product companies to rely on Agile methodologies to shrink their release cycles to days and weeks. SaaS and Cloud bring complexity and sophistication in the software architecture which makes product comprehension more time-consuming and harder. 

\textit{System models can help comprehend software systems better and faster.} In his seminal book ”Mental Models”, Johnson-Laird says, ”Human beings understand the world by constructing models of it in their minds.”. In "Surely you're joking Mr. Feynman" \cite{feynman1992surely}, Feynman describes his approach to understanding something new as making up examples and constructing an object such that it fits all the conditions, thus creating a model of the system and refining it based on new information to get a better understanding. We view software system comprehension as a constructivist learning process that relies on the mental models the learners create and refine as they interact with the software system. Our idea is to augment the mental model creation and refinement through the use of a lightweight system model. This paper presents an idea of using a modeling language and a learning approach to do so. 

\textit{Personal vs. general model}. It is important to be clear of the role these models play. They are used to help refine the mental model of the learner. Given that the mental models are a function of the existing knowledge, the system models will be a function of the existing knowledge as well. As understanding improves, the models will become more refined. These models shouldn't be thought of as general models for the system that can be used by anyone else as well. They are personal also in the sense that they use the vocabulary the learner is comfortable with and hence is best understood by learners themselves. These models can serve other purposes too - for learner to communicate their understanding crisply, or to create a repository of personal models for others to jump-start other learners with similar existing knowledge; however, these usages are beyond the scope of this paper.

Section \ref{sec:relatedwork} presents the related work in the areas of mental models and constructivist learning, as well as how system models can aid learning. Section \ref{sec:motivation} presents the motivation for using transition system models to aid comprehension and describes the modeling language we propose to use. Section \ref{sec:usingmodeling} takes a simple system example and describes the learning approach that enables comprehension for the learner. Section\ref{sec:future} outlines the areas in which future work is planned. 

\section{Related Work}
\label{sec:relatedwork}

\textit{Constructivist View of Learning.} 
According to constructivist view of learning \cite{dennick2016constructivism}, \cite{amineh2015review}, each learner constructs new knowledge for themselves based on their existing knowledge and they play an active role in their learning. Dennick \cite{dennick2016constructivism} suggests that the constructivist view aligns with what the brain does: attempt to extract meaning from the world by interpreting experience through existing knowledge and then building and elaborating new knowledge. Sterman \cite{sterman1994learning} suggests that active modeling occurs well before sensory information reaches the areas of the brain responsible for conscious thought. Thus, to understand something well, it is important for the learner to be actively participating in the learning process through interactions and constructing their own knowledge. Building mental models is the primary way of doing this.

\textit{Mental Models}
Jonassen et.al. \cite{jonassen1999mental} provides a working definition of a mental model in the context of constructivist learning environments that consists of: a) an awareness of the structural components of the system and their descriptions and functions, b) knowledge of the structural inter-relatedness of those components, c) a causal model describing and predicting the performance of the system, and d) a runnable model of how the system functions. Westbrook \cite{westbrook2006mental} defines a mental model as having (1) key components, (2) relationships between those components, and (3) techniques for interacting with the system or process. These closely map to our definition of software system comprehension. Good mental models form through engagement, interactivity and multi-modal content \cite{rapp2005mental}, all of which are expected to be present in some form in a typical onboarding process. Enabling novice engineers to build and refine mental models during their onboarding is likely to enable better and faster comprehension. 

\textit{Models aid systems understanding}
We are specifically interested in understanding of systems. From system dynamics perspective, a model is a tangible aid to imagination and learning, a transitional object to help people make better sense of a partly understood world \cite{morecroft2004mental}. Giving an example of child's play, Morecraft et al. talk about how the activity of repeatedly playing with a toy (a gear set for example) leads to a much more sophisticated understanding and deeper appreciation of the dynamics of the toy because the mental model goes through a series of "transitions" from naive to more sophisticated through repeated use of the toy as a transitional object. Feedback about the real world alters our mental model (Sterman \cite{sterman1994learning}) and enables double-loop learning (Argyris \cite{argyris1991teaching}. Thus having a system model alongside the mental model helps refine the mental models and aids the understanding significantly. Zeigler presents DEVS (Discrete Events) system specification as a seven tuple \cite{zeigler2000theory} to capture the dynamics of the system which is a simple model and lends itself well to reasoning. Lee and Seshia have used a similar model for discrete dynamics modeling of embedded systems. These modeling approaches derive from systems theory and Mealy \cite{mealy1955method} and Moore \cite{moore1956gedanken} machines (finite-state machines for discrete systems). Choppella et al. \cite{choppella2021algodynamics} use transition systems to model interactive algorithms. We want to similarly extend these models and use transition systems as a modeling language for the novice engineers. 

\section{Transition system as a modeling language}
\label{sec:motivation}

As Jonassen et.al. \cite{jonassen1999mental} suggest, to claim someone has understood a software system, they should have a robust mental model using which they can answer questions and make decisions. A formal model can be used as a transitional object \cite{morecroft2004mental} to iterate over mental models. We propose system models built using transition systems as a way to enhance software system comprehension; these are focused on dynamics of the system, use mathematical notations to allow reasoning with the model, and are easy to learn and use (have limited vocabulary).

\subsection{Transition system definition}
A transition system is a six tuple $\{X,X^0,U,f,Y,h\}$ where $X$ is the set of states, $X^0$ is the set of initial states, $U$ is the set of actions, $f$ is the transition relation which is a subset of $(X\times U)\times X$, $Y$ is the set of observables (or output space), and $h$ is a {\em display map\/}, mapping states to observables, $y$ = $h(x)$. 

\begin{figure}[ht]
\centering
\includegraphics[scale=1]{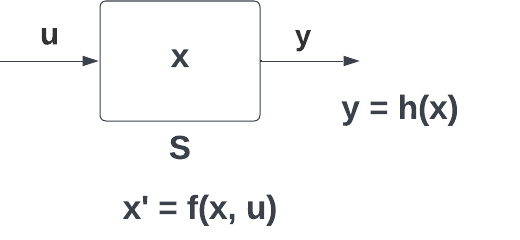}
\caption{Transition system S (x: state, u: action, y: observable, h: display map, f: transition relation)}
\label{fig:ts}
\end{figure}

The transition system models the dynamics of the system as depicted in Figure \ref{fig:ts}.

Let's see how a simple system like a traffic light can be modeled using a transition system. For this, we first need to know what behavior we want to model. 

\textit{Desired Behavior:} Traffic lights flip (green $\rightarrow$ yellow $\rightarrow$ red and reverse) on pre-defined interval - every transition between Red and Green will go through Yellow. There is a manual switch available that can switch the traffic light off (usually for traffic police to take over) and the light is (considered) Black.  

\textit{Transition system model:} For the six tuple $\{X,X^0,U,f,Y,h\}$ for this system, 
$X=\{Red, Yellow, Green, Black\}$, 
$X^0 = \{Black\}$, 
$U=\{timerflip, manualswitch\}$, 
$Y$ = $X$, and 
$h$ is an identity function. 
\begin{description}
    \item $f(u,x)$
    \item $\quad=Yellow$ if $x \in \{Red\}$, $u \in \{timerflip\}$
    \item $\quad=Green$ if $x \in \{Yellow\}$, $u \in \{timerflip\}$
    \item $\quad=Red$ if $x \in \{Green\}$, $u \in \{timerflip\}$
    \item $\quad=Red$ if $x \in \{Black\}$, $u \in \{timerflip\}$
    \item $\quad=Black$ if $x \notin \{Black\} $, $u \in \{manualswitch\}$
\end{description}









\section{Applying system modelling to comprehending software systems}
\label{sec:usingmodeling}
To illustrate how transition system models can be used to comprehend better, we use a case study.

TD is a SaaS product startup that sells a Todo list management software myTodo. myTodo has thousands of users and there is a long list of enhancement requests that the team at TD is working on. Shanks has been hired recently from campus and he is ramping up on the product knowledge so that he can quickly start building product enhancements. 

myTodo is a simple todo list management software. A user can add todos, remove a todo, or mark them as done. A todo can have a due date and the system marks it as delayed if it is not marked done within the due date and triggers a notification to the user.    

Shanks has the following information and resources after first week of his onboarding:
\begin{itemize}
    \item Key use cases and scenarios the product is used in, and access to old feature specs and customer help documentation. 
    \item Product architecture and component overview (1 hour live session) 
    \item Walk through the code organization, access to source code and the test cases for the product. 
    \item A walk through of the live product and access to a test site
    \item Introduction to key subject matter experts (SMEs) who are available to answer specific product questions 
\end{itemize}

After basic understanding of the product and key use cases, Shanks should start building the first version of the model of this software system. 
\subsection{Creating the first cut of the model}
Assuming Shanks understands how to build a transition system model for simple systems, he starts building the six-tuple $\{X,X^0,U,f,Y,h\}$ of the transition system through the steps mentioned below. Note that he can go back and forth on these steps as understanding improves, these do not need to be followed linearly. 
\begin{enumerate}
    \item Identify various inputs to the system and create a list of actions supported by the system. Shanks identifies these actions: 1) add a todo, 2) remove a todo, and 3) mark a todo as done.     
    \item Identify what the system presents as output for various actions and create a list of observables. List of todos and the todo on which the last action was performed seem to be important information for user, so $y=(l,t)$ is selected.   
    \item Using $Y$ and $U$ and the current understanding of the system behavior under different actions, try to identify the states the system should traverse and the information it should hold. This is the hardest part of this exercise. In this case, todo completion is considered an important event, so Shanks chooses these states: 1) No todos in the list, 2) Some todos that are not done (and hence notifications can be generated), and 3) All todos are marked done. Hence $x \in X$ can be thought of as a three-tuple $(s,l, t)$, where s is the name of the state, l is the todo list, and t is the last todo which was acted upon.  
    \item Define the first cut of display map $h$ based on $X$ and $Y$. This is straightforward given we defined the information to be held in $X$ using what we need in $Y$. In this case, this will be a simple function that extracts $l$ and $t$ from $x$ to get $y$. 
    \item Define the set of initial states $X^0$ based on the what is observed in the test site. In this case, $X^0$ will be a state where the list is empty. 
    \item Using $U$, $X$, and the behavior observed (as well as understanding from other sources), define the transition relation ($f$). This may also influence the set of information required to be held in the state, so update $X$ as appropriate.
\end{enumerate}

\subsection{Transition system model of myTodo}
Here is the transition system for the software product Shanks is trying to understand.  

The system has three states: 1) there are no todo items ($NoItems$), 2) there is at least one in-progress todo items ($SomeInProgressItems$), or 3) every todo item is marked done ($AllDoneItems$). Initial state has no todo items ($NoItems$). Three actions are possible: add a new todo item ($Add$), remove a todo item ($Remove$), or mark an existing todo item as done ($MarkDone$). Output $Y$ contains the list of todos ($l$) and the todo that was acted upon most recently ($t$) and this is used by the UI to render various todos differently based on their status (notdone, done or delayed).  

A transition system to capture the dynamics of this system is a six tuple $\{X,X^0,U,f,Y,h\}$, where 
\begin{itemize}
    \item $x = (s,l,t) \in X$, $s$ is the state name , $l$ is the list of todos, and $t$ is the todo which was acted upon most recently. $s \in {\textbf{N}oItems,\textbf{S}omeInProgressItems,\textbf{A}llDoneItems}$ and $l = list(t_1,t_2,..)$. For brevity, we will represent state names by their first letters. $s$ is of type String, $l$ is of type objectid list, and $t$ is of type objectid. We use 0 as an invalid/unavailable id reference.   
    \item $X^0 = \{(N, list(),0\}$, 
    \item $u = (e,t) \in U$, $e$ is the name of the action (or event) and $t$ is the task the action refers to. $e \in \{\textbf{A}dd, \textbf{R}emove, \textbf{M}arkDone\}$ and $t \in \{t_1,t_2,.. \}$. For brevity, we will represent event names by their first letters. $e$ is of type String and $t$ is of type objectid.
    \item $y = (l,t) \in Y$, $l$ is the list of todos, $t$ is the todo which was acted upon most recently.
    \item $h(x) = y$ where $x=(s,l,t)$ and $y = (l,t)$. 
    \item $f(u,x)=x'$ can be represented in terms of its components. $(s',l',t_{last}') = f((e,t),(s,l,t_{last}))$ such that \\
    \begin{description}
        \item $(s',l',t'_{last}) $
        \item $\quad=(S, l+\{t\},t)$ if $e = A$
        \item $\quad=(S, l-\{t\},t)$ if $u = R,\quad   s \notin \{N\}, |l| > 1$
        \item $\quad=(N, l-\{t\},t)$ if $u = R,\quad  s \notin \{N\},|l| = 1$
        \item $\quad=(A, l-\{t\},t)$ if $u = R,status(t) \notin {done} \quad  s \notin \{N\},|l| > 1, |inprogress(l)| = 1$
        \item $\quad=(S, l,t)$ if $u = M,\quad s \in \{S\},|inprogress(l)| > 1$
        \item $\quad=(A, l,t)$ if $u = M,\quad s \in \{S\},|inprogress(l)| = 1$ 
    \end{description}
\end{itemize}
\textit{$|inprogress()|$ returns the count of todos with status $\neq$ done. $status()$ returns the status of the todo (notdone, done, or delayed). } 
\subsection{Refining the model based on interactions and feedback}
Once the first cut of the model is available, Shanks should interact with it in three ways: 1) Generate questions based on the model and ask the subject matter experts, 2) Simulate the model to predict the system behavior and compare it with actual system behavior, and 3) Map the transition function logic components to their implementation

Here are two examples of these interactions and how the received feedback can enhance the model and understanding: 
\begin{itemize}
    \item \textit{Question}: How does the system handle due date expiry? The answer to this can help Shanks model the expiry better, maybe as just another event, thus extending $U$ and updating transition function $f$ to handle this event for all states. 
    \item \textit{Prediction}: Model predicts that when a delayed todo is removed, it is removed normally. Shanks may notice in actual system that a delayed todo can't be removed. Shanks may discover that delayed todo can be marked as done only by the user's manager! This exposes him to an entirely new area of access control and user relationship in the product, thus extending his model and understanding.   
\end{itemize}

As mentioned before, models benefit from interactions and feedback. The learner should create opportunities for interactions: talk to customer support teams, attend team meetings, read and discuss product documents, etc. This learning approach is an iterative process. The updated model (refined based on the feedback) is used to perform the above three activities again and another round of learning cycle is triggered. These learning cycles should continue till the learner feels confident about their system comprehension level. 

More (or less) states are possible in the first cut of the model - For example, a list with delayed todos can be a different state than when none exist. Initial understanding of the learner drives this choice. Learning cycles will ensure that the model gets closer to the real system, but never too close to become hard for the learner to handle; abstraction is key to modeling. 
\section{Limitations and Future Work}
\label{sec:future}
This paper presents initial idea of a personal modeling language and approach for novice engineers to aid their comprehension of software systems and has some limitations. As part of future work, we intend to work on a few specific areas.

\textit{Complex systems with large number of states}. We used a simple system for the illustration of the modeling language and the approach which has only a few states. For a complex system, we do not expect that the model will have large number of states since abstraction level need to match the current comprehension level (mental model). However, as comprehension improves, the model can be refined to a different abstraction level as required by the learner. As part of our future work, we plan to construct examples for reasonably complex open-source systems and show hierarchies of such models mapping to different comprehension levels. 

\textit{A large system consisting of multiple sub-systems}. A large system can be considered to be composed of multiple interconnected systems. Each system can be understood through the approach described above, and the interconnection can be modeled and understood by focusing on $U$ and $Y$ of each component and of the system as a whole, similar to Classic DEVS Coupled Models by Zeigler \cite{zeigler2000theory}.We will formalize the interconnect modeling as part of our future work. 

\textit{Effectiveness of the approach}. We have taught this as part of a course for a small cohort of final-year students. We are now working to design the pedagogy and content for a workshop module and a full course, to be used with novice engineers and final-year students. Our goal will be to study the effectiveness of the model and approach, as well as the pedagogy and content. This is the key future work that we have started working on.

\bibliographystyle{IEEEtran}
\bibliography{icse-nier-2023}

\begin{thebibliography}{10}
\providecommand{\url}[1]{#1}
\csname url@samestyle\endcsname
\providecommand{\newblock}{\relax}
\providecommand{\bibinfo}[2]{#2}
\providecommand{\BIBentrySTDinterwordspacing}{\spaceskip=0pt\relax}
\providecommand{\BIBentryALTinterwordstretchfactor}{4}
\providecommand{\BIBentryALTinterwordspacing}{\spaceskip=\fontdimen2\font plus
\BIBentryALTinterwordstretchfactor\fontdimen3\font minus
  \fontdimen4\font\relax}
\providecommand{\BIBforeignlanguage}[2]{{%
\expandafter\ifx\csname l@#1\endcsname\relax
\typeout{** WARNING: IEEEtran.bst: No hyphenation pattern has been}%
\typeout{** loaded for the language `#1'. Using the pattern for}%
\typeout{** the default language instead.}%
\else
\language=\csname l@#1\endcsname
\fi
#2}}
\providecommand{\BIBdecl}{\relax}
\BIBdecl

\bibitem{sim1998ramp}
S.~E. Sim and R.~C. Holt, ``The ramp-up problem in software projects: A case
  study of how software immigrants naturalize,'' in \emph{Proceedings of the
  20th international conference on Software engineering}.\hskip 1em plus 0.5em
  minus 0.4em\relax IEEE, 1998, pp. 361--370.

\bibitem{10.1145/1882291.1882313}
\BIBentryALTinterwordspacing
M.~Zhou and A.~Mockus, ``Developer fluency: Achieving true mastery in software
  projects,'' in \emph{Proceedings of the Eighteenth ACM SIGSOFT International
  Symposium on Foundations of Software Engineering}, ser. FSE '10.\hskip 1em
  plus 0.5em minus 0.4em\relax New York, NY, USA: Association for Computing
  Machinery, 2010, p. 137–146. [Online]. Available:
  \url{https://doi.org/10.1145/1882291.1882313}
\BIBentrySTDinterwordspacing

\bibitem{ju2021case}
A.~Ju, H.~Sajnani, S.~Kelly, and K.~Herzig, ``A case study of onboarding in
  software teams: Tasks and strategies,'' in \emph{2021 IEEE/ACM 43rd
  International Conference on Software Engineering (ICSE)}.\hskip 1em plus
  0.5em minus 0.4em\relax IEEE, 2021, pp. 613--623.

\bibitem{johnson2010learning}
M.~Johnson and M.~Senges, ``Learning to be a programmer in a complex
  organization: A case study on practice-based learning during the onboarding
  process at {G}oogle,'' \emph{Journal of Workplace Learning}, 2010.

\bibitem{dagenais2010moving}
B.~Dagenais, H.~Ossher, R.~K. Bellamy, M.~P. Robillard, and J.~P. De~Vries,
  ``Moving into a new software project landscape,'' in \emph{Proceedings of the
  32nd ACM/IEEE International Conference on Software Engineering-Volume 1},
  2010, pp. 275--284.

\bibitem{10.1145/1721654.1721672}
\BIBentryALTinterwordspacing
M.~Armbrust, A.~Fox, R.~Griffith, A.~D. Joseph, R.~Katz, A.~Konwinski, G.~Lee,
  D.~Patterson, A.~Rabkin, I.~Stoica, and M.~Zaharia, ``A view of cloud
  computing,'' \emph{Commun. ACM}, vol.~53, no.~4, p. 50–58, apr 2010.
  [Online]. Available: \url{https://doi.org/10.1145/1721654.1721672}
\BIBentrySTDinterwordspacing

\bibitem{feynman1992surely}
\BIBentryALTinterwordspacing
R.~Feynman, R.~Leighton, and E.~Hutchings, \emph{"Surely You're Joking, Mr.
  Feynman!": Adventures of a Curious Character}.\hskip 1em plus 0.5em minus
  0.4em\relax Vintage, 1992. [Online]. Available:
  \url{https://books.google.co.in/books?id=N3eLdCTynR0C}
\BIBentrySTDinterwordspacing

\bibitem{dennick2016constructivism}
R.~Dennick, ``Constructivism: reflections on twenty five years teaching the
  constructivist approach in medical education,'' \emph{International journal
  of medical education}, vol.~7, p. 200, 2016.

\bibitem{amineh2015review}
R.~J. Amineh and H.~D. Asl, ``Review of constructivism and social
  constructivism,'' \emph{Journal of Social Sciences, Literature and
  Languages}, vol.~1, no.~1, pp. 9--16, 2015.

\bibitem{sterman1994learning}
J.~D. Sterman, ``Learning in and about complex systems,'' \emph{System dynamics
  review}, vol.~10, no. 2-3, pp. 291--330, 1994.

\bibitem{jonassen1999mental}
D.~H. Jonassen and P.~Henning, ``Mental models: Knowledge in the head and
  knowledge in the world,'' \emph{Educational technology}, pp. 37--42, 1999.

\bibitem{westbrook2006mental}
L.~Westbrook, ``Mental models: a theoretical overview and preliminary study,''
  \emph{Journal of Information Science}, vol.~32, no.~6, pp. 563--579, 2006.

\bibitem{rapp2005mental}
D.~N. Rapp, ``Mental models: Theoretical issues for visualizations in science
  education,'' in \emph{Visualization in science education}.\hskip 1em plus
  0.5em minus 0.4em\relax Springer, 2005, pp. 43--60.

\bibitem{morecroft2004mental}
J.~Morecroft, ``Mental models and learning in system dynamics practice,''
  \emph{Systems modelling: Theory and practice}, pp. 101--126, 2004.

\bibitem{argyris1991teaching}
C.~Argyris, ``Teaching smart people how to learn,'' \emph{Harvard business
  review}, vol.~69, no.~3, 1991.

\bibitem{zeigler2000theory}
B.~P. Zeigler, T.~G. Kim, and H.~Praehofer, \emph{Theory of modeling and
  simulation}.\hskip 1em plus 0.5em minus 0.4em\relax Academic press, 2000.

\bibitem{mealy1955method}
G.~H. Mealy, ``A method for synthesizing sequential circuits,'' \emph{The Bell
  System Technical Journal}, vol.~34, no.~5, pp. 1045--1079, 1955.

\bibitem{moore1956gedanken}
E.~F. Moore \emph{et~al.}, ``Gedanken-experiments on sequential machines,''
  \emph{Automata studies}, vol.~34, pp. 129--153, 1956.

\bibitem{choppella2021algodynamics}
V.~Choppella, K.~Viswanath, and M.~Kumar, ``Algodynamics: Algorithms as
  systems,'' in \emph{2021 IEEE Frontiers in Education Conference (FIE)}.\hskip
  1em plus 0.5em minus 0.4em\relax IEEE, 2021, pp. 1--9.

\end{thebibliography}

\end{document}